\documentstyle[12pt]{article}

\begin{document}

\baselineskip 4ex
\parindent 2em

\centerline{ \bf {Fundamental properties of a vortex  } }
\centerline{ \bf {in a d-wave superconductor} }
\vskip 1.5cm
\centerline{M. Ichioka$^1$, N. Hayashi$^{\dagger}$$^2$, N.
Enomoto$^{\dagger}$$^3$ and K. Machida$^{\dagger}$$^4$}

\vskip 1.5cm

\centerline{ Department of Physics, Kyoto University, Kyoto
606-01, Japan}

\centerline{ $^{\dagger}$Department of Physics, Okayama University, Okayama
700, Japan}

\vskip 2cm

\noindent
The vortex  core structure in a d-wave superconductor
is analyzed on the basis of the quasi-classical Eilenberger theory beyond
the Ginzburg-Landau framework.
The current and magnetic field distributions around an isolated vortex
break circular symmetry seen in s-wave pairing  and show four-fold
symmetry, reflecting the internal degrees of freedom in  d-wave pairing,
{\it i.e.} $\hat k_x^2-\hat k_y^2$ in reciprocal space through the low
lying quasi-particle excitations.
The peculiar orientation of the flux line lattice observed recently in a
cuprate is argued in light of the present theory.


\vskip 0.5cm
\noindent
$^1$oka@ton.scphys.kyoto-u.ac.jp

\noindent
$^2$hayashi@mp.okayama-u.ac.jp

\noindent
$^3$enomoto@mp.okayama-u.ac.jp

\noindent
$^4$machida@mp.okayama-u.ac.jp

\newpage

Much attention has been focused on the high temperature
cuprate superconductors. Although precise pairing
symmetry has not been identified yet, it is
recognized\cite{review} that d$_{x^2-y^2}$-wave
symmetry is most probable.

Recently, Keimer {\it et.al.}\cite{keimer} reported the small
angle neutron scattering
experiment on
YBa$_2$Cu$_3$O$_7$
to see the flux line lattice (FLL) in a magnetic field
region; $0.5{\rm T}\leq H \leq 5{\rm T}$ and found that the vortices form an
oblique lattice with two equal lattice constants and an angle of 73$^\circ$
between the two primitive vectors. FLL is oriented
such that the nearest neighbor direction of vortices
makes 45$^\circ$ from the $a$-axis (or $b$-axis) of the
underlying orthorhombic crystal lattice when the
field $H$ is applied to the $c$-axis. This result is highly
non-trivial to be understood within the framework of the Ginzburg-Landau
theory (GL). In order to stabilize such an
orientation of FLL, one of the primitive vectors must be aligned, making an
angle of 9$^\circ$ from the $a$-axis.
This means that we need the orientational energy of a higher order form
$\Delta_0^n\cos n\theta$ with $n=10$ in the GL free
energy,  where $\theta$ is the angle between one of the
primitive vectors and the $a$-axis (or $b$-axis). This is quite unlikely in
the spirit of the order parameter ($\Delta_0$) expansion.\cite{takanaka}

Within the framework of GL, which is widely and
conveniently used to discuss electromagnetic properties of cuprates, s-wave
and d-wave superconductivity are not
distinguished as far as the order parameter is single component because
only the number of the order
parameter components are relevant to construct the GL
free energy. This is particularly true for the vortex structure. Therefore,
the vortex core structure of d-wave
pairing has been considered to be  cylindrically symmetric with a
singularity at the center,
as in s-wave superconductors,  within GL framework. The supercurrent and
magnetic field distributions around an isolated vortex line are also
cylindrically symmetric unless the crystalline anisotropy is implemented.
(Note that the Cu-O
plane in cuprates is quite isotropic in this respect.)

Motivated by the neutron experiment and the above
considerations, we undertake to examine the vortex core structure of d-wave
superconductivity. This study underlies the
fundamental understanding of electromagnetic properties of cuprates. We
hope that the internal degrees of freedom
of a d-wave pair characterized by the relative coordinates such as
$\Delta(\vec k)\propto {\hat k}^2_x-{\hat k}^2_y$ might affect on
determining properties related to spatial
variations which are governed by the center of mass coordinates of a pair,
such as a core structure or interface\cite{shiba} where the pair function
is forced to
vary spatially.

The Eilenberger theory\cite{eilenberger} is best suited for
our purposes. Because this quasi-classical approach
was originally designed to describe the conventional s-wave
superconductors beyond the GL framework microscopically.
This method easily allows us to treat the two different coordinates systems
mentioned above and to discuss the physics of the length scale $\sim \xi_0$
(the coherence length), ignoring the shorter scale $k_F^{-1}$ ($k_F$; Fermi
wave number).

The purpose of this paper is to analyze the vortex
structure for a d-wave pairing  beyond the GL framework, namely, within the
quasi-classical Eilenberger theory. We will conclude that reflecting the
internal degrees of freedom of a d-wave pairing function,
the vortex core structure is anisotropic and the cylindrical symmetry
around a vortex line seen in s-wave
pairing is spontaneously broken. The associated supercurrent distribution
and magnetic field distribution also become four-fold symmetric. This
spontaneous symmetry
breaking of the vortex structure is generic in anisotropic superconductors.
Other non-trivial pairing states which are known to
exist in heavy Fermion superconductors such as UPt$_3$\cite{machida} and
superfluid $^3$He\cite{volovik}
must give rise to own particular anisotropic vortex core
structures, corresponding to their particular symmetries both of orbital
and spin structures  of the pairing
function. The cylindrical symmetry around a vortex line is realized only
when the pairing is s-wave.

We starts with a transport-like
Eilenberger equation for the
quasi-classical Green
function $\hat{g}({\vec r},
\theta,i\omega_n)=-i\pi\pmatrix{g & if\cr -if^{\dagger} &-g \cr}$
in a 2$\times$2 matrix form, namely,

$$i\vec{v}_F\cdot\nabla\hat{g}({\vec r}, \theta,i\omega_n)
+[\pmatrix{i\omega_n & -\Delta(\vec{r},\theta)\cr \Delta^*(\vec{r},\theta)
&-i\omega_n\cr}, \  \ \hat{g}({\vec r}, \theta,i\omega_n)]=0$$

\noindent
supplemented by the normalization condition: $\hat{g}({\vec r},
\theta,i\omega_n)\cdot\hat{g}({\vec r},
\theta,i\omega_n)=-\pi^2\cdot\hat{1}$. The Fermi velocity is $v_F$ (The
Fermi surface is assumed to be a circular). The order parameter
$\Delta(\vec{r},\theta)$ is described by
two kinds of the coordinates, namely,
the two-dimensional center of gravity  coordinates
$\vec{r}=(x, y)$ and the relative coordinate $\theta$
which is measured from the $a$-axis (or $x$-axis). The
bracket $[\ \ , \  \ ]$ is a commutator. The self-consistent equation is given

$$\Delta(\vec{r},\theta)=N_02\pi T\sum_{\omega_n>0}\int{d\theta'\over
2\pi}V(\theta,\theta')f({\vec r}, \theta',i\omega_n)$$

\noindent
where the pairing interaction $V(\theta,\theta')$ is assumed to be
separable. The density of states at the Fermi surface is $N_0$.
The associated supercurrent $\vec{j}({\vec r}, \theta)$ is given  in term
of ${g}({\vec r}, \theta,i\omega_n)$ by

$$\vec{j}({\vec r}, \theta)=2ev_FN_02\pi
T\sum_{\omega_n>0}\int{d\theta'\over 2\pi}{\vec{k}\over i}g({\vec r},
\theta',i\omega_n).$$

\noindent
The magnetic field is determined through ${4\pi \over c}{\vec
j}={\vec\nabla} \times{\vec H}.$
Instead of solving the above Eilenberger equation, it is
more convenient to use the following parametrization
devised by Schopohl and Maki\cite{schopohl}

$$\hat{g}=-i\pi{1\over 1+ab}\pmatrix{1-ab&2ia\cr -2ib&-1+ab\cr}.$$

\noindent
The Riccati equation is satisfied by $a(r_{\parallel})$ and $b(r_{\parallel})$:

$$v_F{\partial a(r_{\parallel})\over \partial r_{\parallel}}+
[2\omega_n+\Delta^*(r_{\parallel})a(r_{\parallel})]a(r_{\parallel})-\Delta(r
_{\parallel})=0.$$

\noindent
The other unknown quantity $b(r_{\parallel})$ is related by
$b(r_{\parallel})=-a(-r_{\parallel})e^{-2i\theta}$ by symmetry in the
isolated
vortex case under consideration. Here we have taken the coordinate system:
$\vec {\hat u}=\cos\theta \vec {\hat x}+\sin\theta \vec {\hat y}, \vec
{\hat v}=-\sin\theta \vec {\hat x}+\cos \theta \vec {\hat y}$, thus a point
$\vec{r}=x \hat{x}+y\hat{y}$ is denoted as $\vec{r}=r_{\parallel}\vec
{\hat{u}}+r_
{\perp}\vec {\hat{v}}$.

We take  the following form for the d-wave vortex
structure as
a test potential:

$$\Delta(\vec{r}, \vec{\hat{k}},\theta)=\Delta_0{\rm
tanh}{r\over \xi_0}\cdot e^{i\phi}\cos2\theta$$

\noindent
and $V(\theta, \theta')=g_d\cos2\theta \cos2\theta'$ with
$g_d$ being the coupling constant for the d-wave channel.
The phase factor is given by $e^{i\phi}={{x+iy}\over {\sqrt{x^2+y^2}}}$.
The numerical computation has been done using the standard algorithm
for solving the first order differential equation.

The stereographic view of the current distribution near the vortex core is
shown in Fig.1. It is clear that the
amplitude of
$|j(\vec r)|$ is not circularly symmetric. The breaking of
the cylindrical symmetry is stronger in the core region
while toward the outer region the cylindrical symmetry is recovered
gradually. The four-fold symmetry of $|j(\vec r)|$
arises from the d-wave nature of the relative coordinate of the pairing
function $\Delta(\vec k)\propto \hat  k_x^2-\hat k_y^2$.
The four small peaks around the core move inside toward the core when the
temperature ($T$) is lowered,
indicating that
the core region is narrowed with decreasing $T$
(compare Figs. 1(a) and (b)). This
feature coincides qualitatively with a general rule known for conventional
s-wave superconductors\cite{klein} and
confirms an interpretation that it arises from Fermionic excitations
bounded by the core associated with d-wave pairing by
Volovik\cite{volovik2}.
The current distribution of $|j(\vec r)|$ extents toward
the 45$^{\circ}$ direction from the $x$-axis and its equivalent directions.
This is physically because there
exist the local quasi-particle excitations along those directions which
contribute to inducing the
supercurrent. In  fact the
local density of states of the quasi-particle
excitations\cite{schopohl} has a characteristic four-fold symmetric shape
for low energies ($\leq \Delta_0$) and
almost circular shape for higher energies ($\geq\Delta_0$). The latter is
responsible for almost
circular distribution of $|j(\vec r)|$ in the outer region.
It is known that in the core region physical quantities are dominated by
lower energy excitations\cite{gygi}.

The associated magnetic field distribution $H(\vec r)$ is displayed in
Figs. 2(a) and (b). It is evident from Fig. 2(a)
that $H(\vec r)$ is
also
four-fold symmetric. The field extends along the $x$-axis and its
equivalent directions, rotated by 45$^{\circ}$ from the
current distribution. This is because the field is much screened by the
induced current along the 45$^{\circ}$ direction, reflecting that the local
quasi-particle excitations with lower energies deviates from the circular
shape. The field distribution becomes almost circular
beyond a characteristic length scale $\xi_0$.
We show a stereographic view of $H(\vec r)$ in Fig. 2(b) where the ridges
are barely seen along the $x$-axis and
its equivalent directions.

While our calculation is based on the test potential as a first step, the
feature of our results reflects qualitatively the essential nature of the
d-wave pairing.  The quantitative aspect might be modified by performing
the self-consistent calculation.
This is particularly true for the degree of the anisotropy in $H(\vec r)$.
In fact, we
can certainly anticipate that the anisotropy in $H(\vec r)$ here is further
amplified by a self-consistent computation.
Because the resulting order parameter profile
around the core, calculated
by assuming
a circular symmetric form:
$\Delta(\vec r)\propto \rm{tanh}{r\over \xi}\cdot e^{i\phi}$
as a starting test potential, comes out to be
a anisotropic one
(Note the square shapes near the center
in the contour
plot of Fig. 3). This will
further lead to the
anisotropic current and field distribution. The full self-consistent
computation is left for a future problem.

Finally, we point out a possibility that other components
of the order parameter may be induced around a d-wave vortex: As shown in
Fig. 4 as an example,
the s-wave
component evaluated
by
$\Delta_s(\vec r)=N_0g_s2\pi T\sum\int{d\theta'\over 2\pi}f(\vec r,
\theta',i\omega_n)$
is induced only in the core region of a
d-wave superconductor
if the pairing interaction has
a small s-wave component $g_s$.
Associated with the spatial inhomogeneity of
the order parameter at a vortex or interface, other components may be
induced. This tendency is in agreement with the similar quasi-classical
analysis of interfaces\cite{shiba} and with the GL theory analysis of
vortices\cite{berlinsky}. The mixing of two kind of order
parameters, e.g. $s+id$-pairing vortex further breaks spatial symmetry,
giving rise to the two-fold symmetric local density of states.

Based on our calculations, let us argue how the flux
line lattice (FLL) differs from the conventional superconductors: The FLL
in a d-wave could be different from the regular triangular lattice in a
s-wave because the latter is led due to the isotropy of the
vortex core. The distorted vortex core structure in the d-wave must be
taken into account when determining the structure of the FLL.

As for the FLL orientation relative to the underlying crystalline lattice,
in the traditional treatment\cite{takanaka} based on
the GL theory the higher order derivatives, or the so-called non-local
corrections are considered to derive an orientational energy. Here the
orientational energy comes from the anisotropic current distribution and
field
distribution. We can speculate a possibility that when placing a nearest
neighbor vortex at a certain distance from the center vortex, it could be
advantageous to place it
where the magnetic field is low to minimize the
condensation energy loss, that is, the 45$^{\circ}$ direction in our case.
This may explain the observation of the neutron experiment\cite{keimer}
mentioned earlier. These arguments should be substantiated  by the detailed
calculations beyond the GL theory.  Here we only show the essence. Together
with these theoretical investigations, the experimental efforts to
understand the fundamental electromagnetic properties of the FLL associated
with a d-wave pairing are
certainly needed, for example, the further small angle neutron scattering
or STM experiments are
desirable to see the actual field distribution, or to directly probe the
current distribution  around a vortex
core, which could distinguish it from s-wave pairing.

We thank N. Schopohl, K. Maki, D. Rainer and K. Takanaka for useful
theoretical information and B. Keimer for discussions on their experiment.

\newpage

{\bf Figure Captions}

\bigskip
\noindent
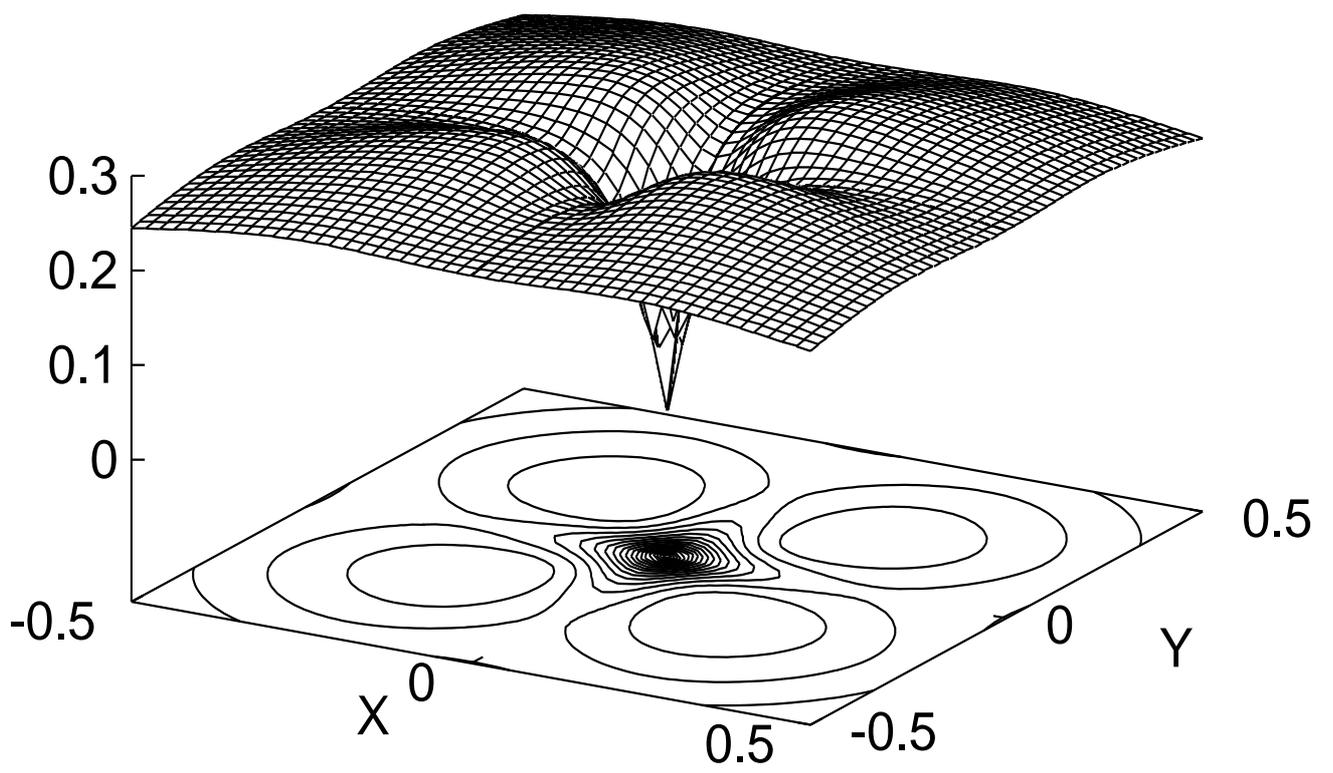
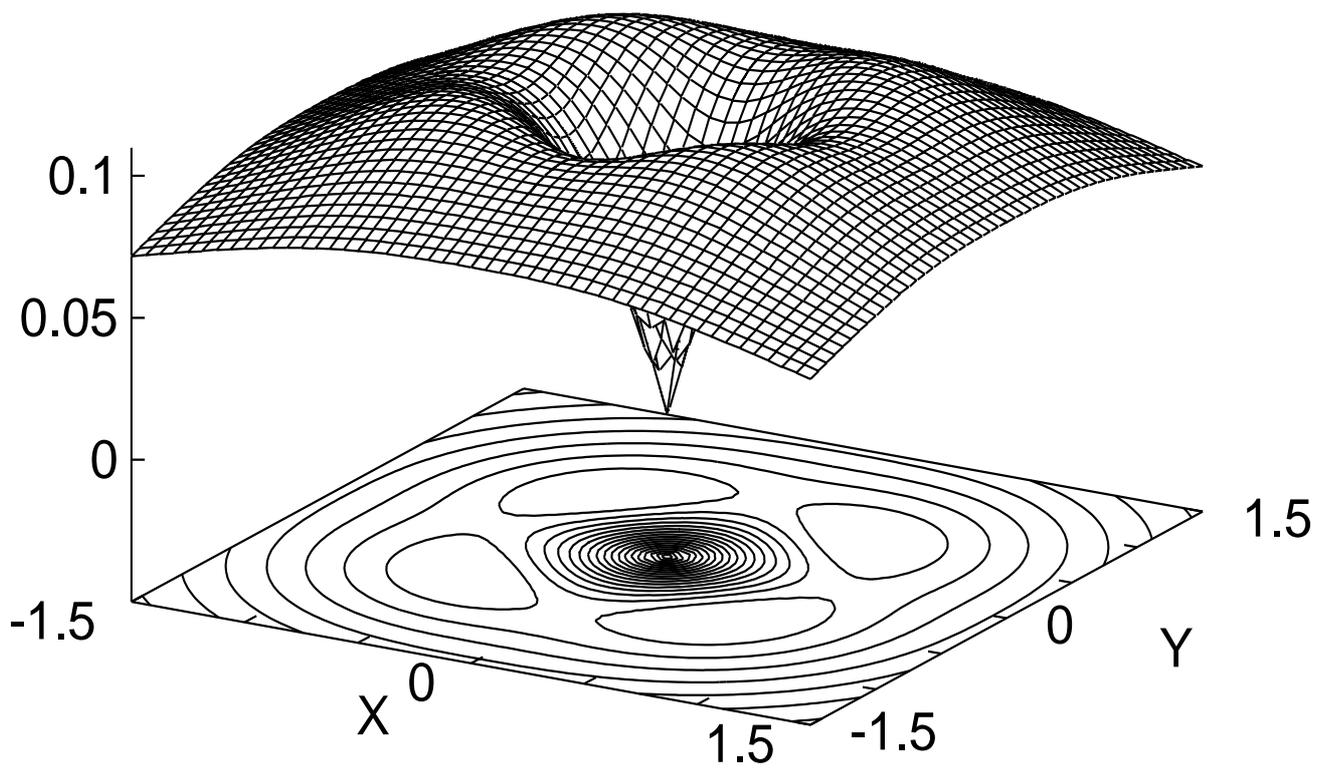
Fig.1: Amplitude of the current density $|\vec{j}(\vec{r})|$ in the core
region at $T/T_c=0.05$ (a) and $T/T_c=0.5$ (b) in the unit of
$-2eN_0v_F\Delta_0$ ($T_c$ is the transition temperature).
There are four small peaks around the core in the 45$^\circ$ and its
equivalent directions. These peaks become clear and move toward the core
upon lowering $T$.
Note the change of the length scale in (a) and (b) which is normalized by
$\xi_0$.

\bigskip
\noindent
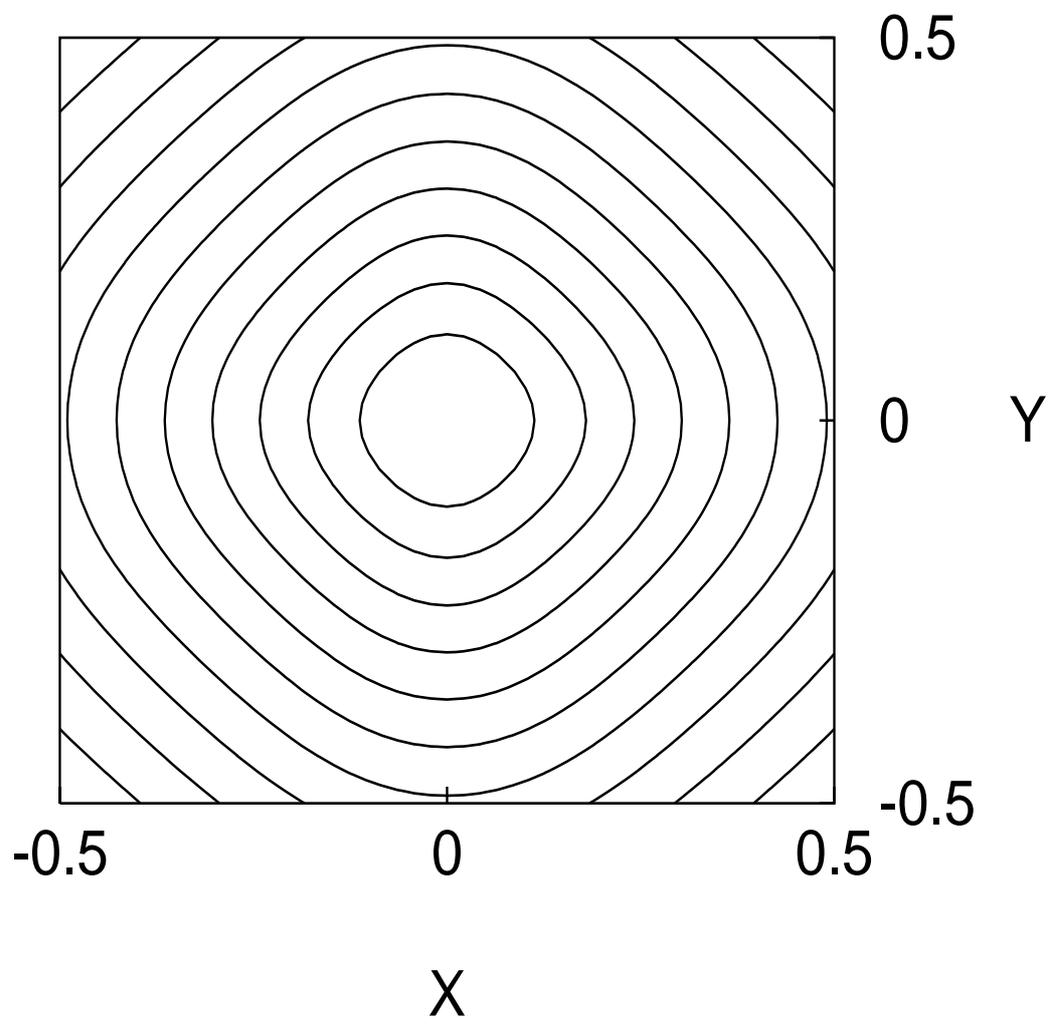
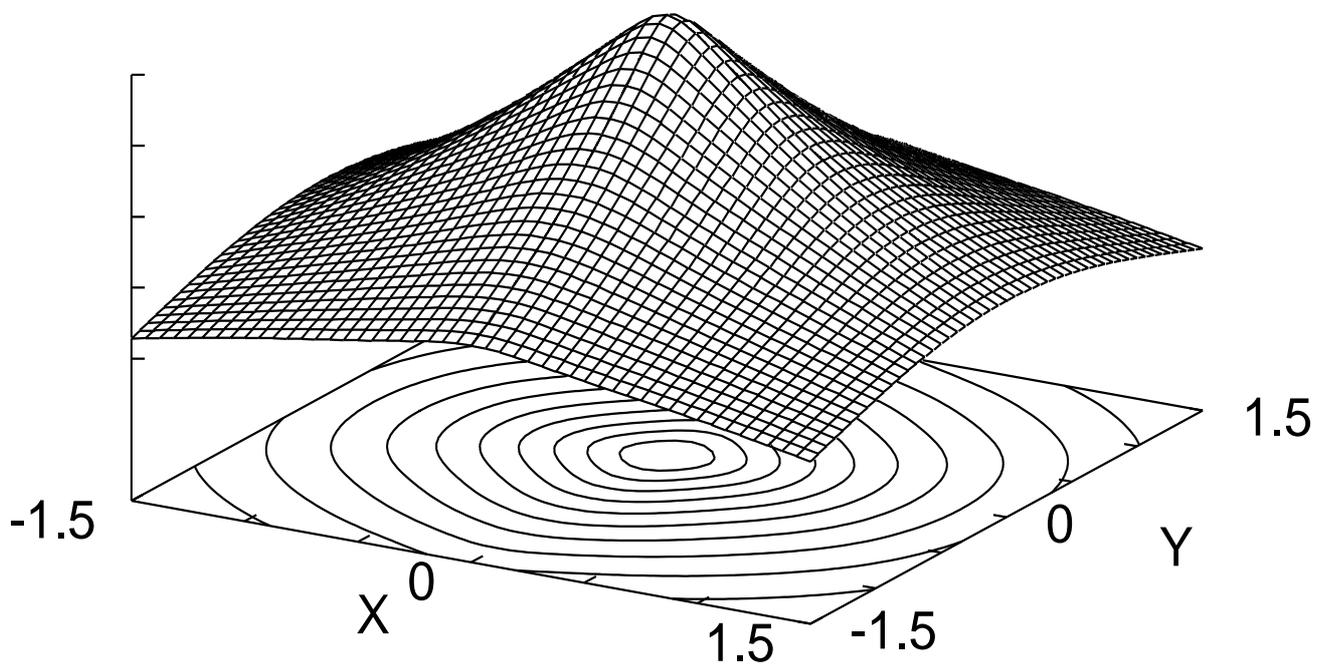
Fig.2: Contour plot (a) and stereographic view (b) of the magnetic field
distribution at $T/T_c=0.05$.
The core region is focused in (a). The field extends along the $x$-axis and
$y$-axis.

\bigskip
\noindent
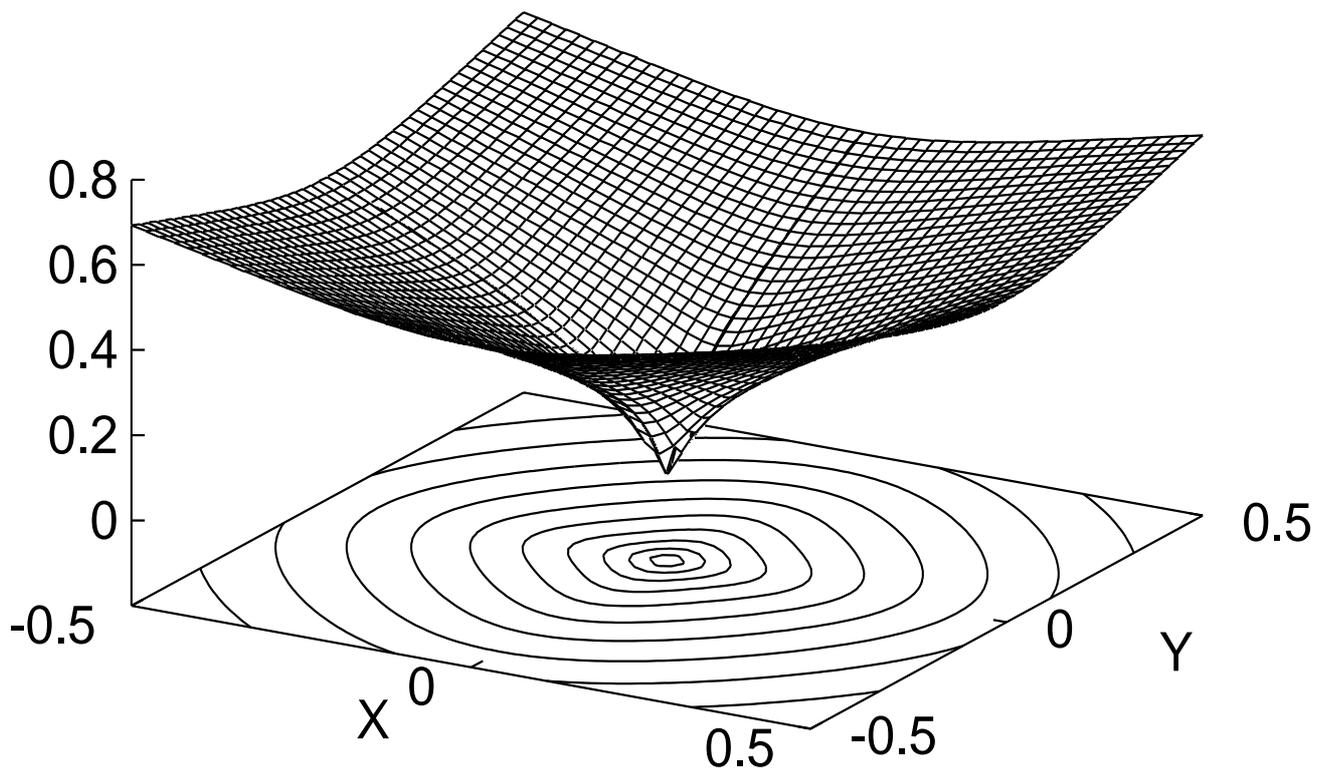
Fig.3: Amplitude of $|\Delta(\vec{r})|$, the $\hat k_x^2 -\hat k_y^2$
component, at $T/T_c=0.05$ calculated by using the resulting Green
function.  It is normalized by $\Delta_0$.
The amplitude is suppressed in the $x$-axis and $y$-axis directions as is
seen from the square shapes in the contour plot.

\bigskip
\noindent
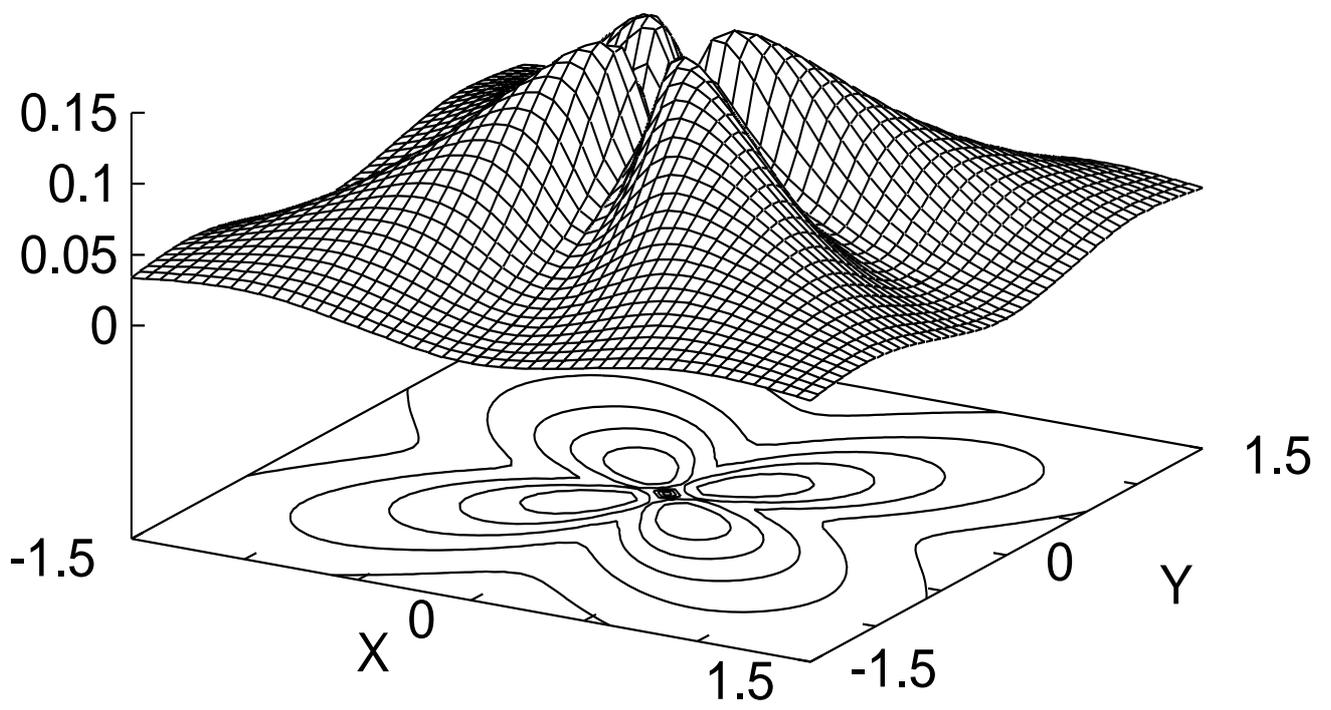
Fig.4: Amplitude of  $|\Delta_s(\vec{r})|/(g_s/g_d)$, the s-wave component,
at $T/T_c=0.05$ calculated by using the resulting Green function. The
s-wave amplitude is induced in the core region.
The contour plot shows that the profile has a four-leafed  clover shape as
coinced with the GL prediction\cite{berlinsky}.


\newpage


\begin{thebibliography}{99}


\bibitem{review} See for example, papers in the Proceedings of M$^2$S
Conference in  Physica {\bf C
235-240}, 1 (1994).



\bibitem{keimer}B. Keimer, {\it et.al.},  Phys. Rev. Lett.
{\bf 73}, 3459 (1994).

\bibitem{takanaka}K. Takanaka, Prog. Theor. Phys. {\bf 50}, 365  (1973). K.
Takanaka and K. Machida, private communication.

\bibitem{shiba} The order parameter mixing is discussed to occur at
interfaces  in M. Matsumoto and H. Shiba, preprint

\bibitem{eilenberger} G. Eilenberger, Z. Phys. {\bf 214}, 195 (1968). See
for review, J. W. Serene and D. Rainer,
Phys. Rep. {\bf 101}, 221 (1983). An analysis of the vortex structure of a
s-wave superconductor is done in L. Kramer and W. Pesch, Z. Phys. {\bf
269}, 59 (1974).

\bibitem{machida} See for example, K. Machida {\it et.al.},
J. Phys. Soc. Jpn. {\bf 64}, 1067 (1995).

\bibitem{volovik} M. M. Salomaa and G. E. Volovik, Rev.
Mod. Phys. {\bf 59}, 533 (1987).

\bibitem{schopohl}N. Schopohl and K. Maki, preprint. K. Maki {\it et.al.},
Physica {\bf B204}, 214 (1995).

\bibitem{klein}U. Klein, Phys. Rev.
{\bf B40}, 6601 (1989) and {\bf B41}, 822 (1989).


\bibitem{volovik2}G. E. Volovik, JETP Lett. {\bf 58}, 455 and 469 (1993)
and Physica {\bf C235-240}, 2411 (1994).

\bibitem{gygi}F. Gygi and M. Schl\"{u}ter, Phys. Rev.
{\bf B43}, 7609 (1991).




\bibitem{berlinsky}P.I. Soininen {\it et.al.}, Phys. Rev.
{\bf B50}, 13883 (1994) and A. J. Berlinsky {\it et.al.}, preprint.
Also see, Y. Ren {\it et.al.}, Phys. Rev. Lett.
{\bf 74}, 3680 (1995).



\end{thebibliography}
\end{document}